\documentclass[fleqn,10pt]{wlscirep}
\usepackage{graphicx}
\usepackage{epstopdf}
\usepackage{color}
\usepackage{physics}
\usepackage{float}

\title{Ising-type Magnetic Anisotropy in CePd$_2$As$_2$}

\author[1,*]{M.\ O.\ Ajeesh}
\author[2]{T.\ Shang}
\author[2]{W.\ B.\ Jiang}
\author[2]{W.\ Xie}
\author[1]{R.\ D.\ dos Reis}
\author[2]{M.\ Smidman}
\author[1]{C.~Geibel}
\author[2]{H.\ Q.\ Yuan}
\author[1]{M.\ Nicklas}

\affil[1]{Max Planck Institute for Chemical Physics of Solids, N\"{o}thnitzer Str.\ 40, 01187 Dresden, Germany}
\affil[2]{Center for Correlated Matter and Department of Physics, Zhejiang University, Hangzhou 310058, China}

\affil[*]{ajeesh@cpfs.mpg.de}

\begin{abstract}

We investigated the anisotropic magnetic properties of CePd$_2$As$_2$ by magnetic, thermal and electrical transport studies. X-ray diffraction confirmed the tetragonal ThCr$_2$Si$_2$-type structure and the high-quality of the single crystals. Magnetisation and magnetic susceptibility data taken along the different crystallographic directions evidence a huge crystalline electric field (CEF) induced Ising-type magneto-crystalline anisotropy with a large $c$-axis moment and a small in-plane moment at low temperature. A detailed CEF analysis based on the magnetic susceptibility data indicates an almost pure $\ket{\pm5/2}$ CEF ground-state doublet with the dominantly $\ket{\pm3/2}$ and the $\ket{\pm1/2}$ doublets at 290~K and 330~K, respectively. At low temperature, we observe a uniaxial antiferromagnetic (AFM) transition at $T_N=14.7$~K with the crystallographic $c$-direction being the magnetic easy-axis. The magnetic entropy gain up to $T_N$ reaches almost $R\ln2$ indicating localised $4f$-electron magnetism without significant Kondo-type interactions. Below $T_N$, the application of a magnetic field along the $c$-axis induces a metamagnetic transition from the AFM to a field-polarised phase at $\mu_0H_{c0}=0.95$~T, exhibiting a text-book example of a spin-flip transition as anticipated for an Ising-type AFM.

\end{abstract}
\begin{document}

\flushbottom
\maketitle

\thispagestyle{empty}

\section*{Introduction}

Materials crystallising in the ThCr$_2$Si$_2$-type structure comprise of such prominent compounds as the first heavy-fermion superconductor CeCu$_2$Si$_2$ \cite{Steglich79} and BaFe$_2$As$_2$, a parent compound to the iron-based high-temperature superconductors \cite{Rotter08}, making them especially attractive for solid-state research in the past decades. The discovery of heavy-fermion superconductivity in  CeCu$_2$Si$_2$ resulted in extensive studies which became crucial for the understanding of unconventional superconductivity. In Ce-based heavy-fermion systems the strength of the hybridisation between the Ce-4\emph{f} electrons and the conduction electrons is particularly important for the physical behaviour at low temperatures. There, the competition between Kondo effect and Ruderman-Kittel-Kasuya-Yoshida (RKKY) interaction along with crystalline electric field (CEF) effects lead to a large variety of different ground-state properties, which might be tuned using external control parameters such as chemical substitution, magnetic field and hydrostatic pressure \cite{Jaccard92, Movshovich96, Grosche96, Yuan03, Lengyel11}.

A large number of the Ce-based compounds crystallising in the ThCr$_2$Si$_2$-type structure order antiferromagnetically at low temperatures. Their magnetism is commonly determined by a large magneto-crystalline anisotropy. This leads to the presence of distinct field-induced metamagnetic transitions \cite{Abe98,Thamizhavel07,Knafo10,Krimmel99,Ota09,Fritsch11,Luo12,Maurya13}. Depending on the strength of the magnetic anisotropy, the nature of the metamagnetic transition(s) may differ. Additionally, the spin structure in the antiferromagnetic (AFM) phase plays an important role in the field-induced metamagnetic transition(s). In this regard, CePd$_2$As$_2$ offers the opportunity to study magnetism in a localised moment antiferromagnet with a huge magneto-crystalline anisotropy.

Recently, the physical properties of polycrystalline CePd$_2$As$_2$, which crystallises in the ThCr$_2$Si$_2$-type structure, were reported~\cite{Shang14}. CePd$_2$As$_2$ undergoes an antiferromagnetic (AFM) ordering at $T_N\approx 15$~K and shows evidence of a metamagnetic transition. However, a detailed investigation on single crystalline samples is necessary in order to understand the magnetic properties. In this work, we report on the magnetic anisotropy of single crystalline CePd$_2$As$_2$. To this end, we carried out magnetic susceptibility, magnetisation, electrical-transport and specific-heat measurements. Our results reveal an Ising-type magnetic anisotropy which accounts for a text-book-like spin-flip metamagnetic transition. The CEF level scheme could be fully resolved based on our experimental data. Furthermore, our analysis suggests a simple, collinear A-type antiferromagnetic spin structure in the AFM state.

\section*{Results}

\subsection*{Magnetic susceptibility and heat capacity}\label{susceptibility}

\begin{figure}[t!]
\centering
\includegraphics[width=0.5\linewidth]{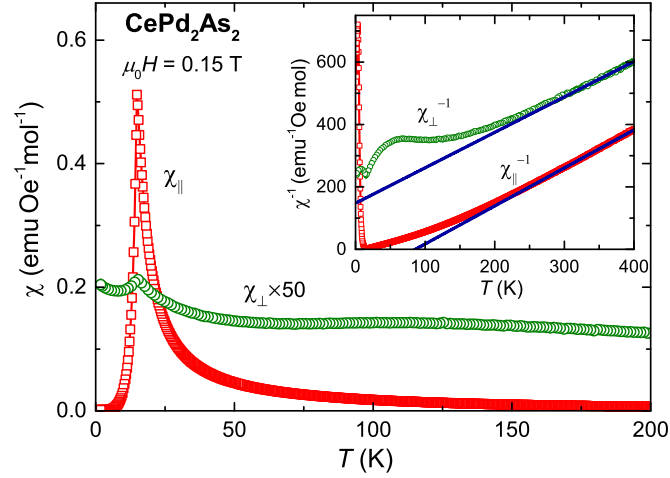}
\caption{Temperature dependence of the magnetic susceptibility $\chi=M/H$ for magnetic fields applied parallel ($\chi_{\|}$) and perpendicular ($\chi_{\perp}$) to the $c$-axis. Here, $M$ is the magnetisation and $H$ is the magnetic field. The inset displays the inverse magnetic susceptibility as a function of temperature. The solid lines represent fits of a Curie-Weiss law, $\chi(T)=C/(T-\theta_{\rm W})$, to $\chi_{\|}(T)$ and $\chi_{\perp}(T)$ in the temperature interval $300{\rm~K}\leq T\leq 400{\rm~K}$ and $350{\rm~K}\leq T\leq 400{\rm~K}$, respectively.}
\label{ChivsT}
\end{figure}

\begin{figure}[t!]
\centering
\includegraphics[width=0.5\linewidth]{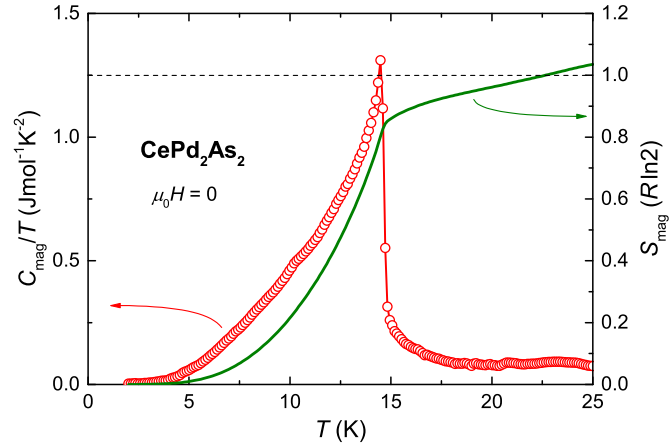}
\caption{Temperature dependence of the magnetic contribution to the specific heat $C_{\rm mag}$ of CePd$_2$As$_2$, plotted as $C_{\rm mag}(T)/T$ (left axis). $C_{\rm mag}$ was estimated by subtracting the specific heat of the non-magnetic reference compound LaPd$_2$As$_2$ from that of CePd$_2$As$_2$. The calculated magnetic entropy $S_{\rm mag}(T)$ is displayed in the unit of $R\ln2$ (right axis).}
\label{HCvsT}
\end{figure}

The temperature dependence of magnetic susceptibility $\chi$ of CePd$_2$As$_2$ with magnetic fields applied parallel ($\chi_{\|}$) and perpendicular ($\chi_{\perp}$) to the crystallographic $c$-axis are depicted in Fig.~\ref{ChivsT}. $\chi(T)$ shows a sharp peak at $T_N=14.7$~K for both field orientations indicating the AFM transition, in good agreement with the results previously reported on polycrystalline samples~\cite{Shang14}. Remarkably, $\chi_{\|}$ is two orders of magnitude larger than $\chi_{\perp}$ implying the presence of a strong magnetic anisotropy. The inverse magnetic susceptibility, $\chi_{\|}^{-1}(T)$ and $\chi_{\perp}^{-1}(T)$ are plotted in the inset of Fig.~\ref{ChivsT}. Above room temperature, the susceptibility data can be fit by a Curie-Weiss law, $\chi(T)=C/(T-\theta_{\rm W})$, where $C$ and $\theta_{\rm W}$ are the Curie constant and the Weiss temperature, respectively. We find $\theta^{\|}_{\rm W}=86~K$ and  $\mu_{\rm eff}= 2.56\mu_{B}$ for $H\parallel c$ and $\theta^{\perp}_{\rm W}= -130~K$  and  $\mu_{\rm eff}= 2.65\mu_{B}$  for $H\perp c$. The obtained effective moments are slightly enhanced compared with the calculated value of $2.54~\mu_{B}$ for a free Ce$^{3+}$ ion. The deviation of $\chi_{\rm \parallel , \perp}(T)$ from a Curie-Weiss law below room temperature can be attributed to CEF effects.

For a free Ce$^{3+}$ ion with total angular momentum $J=5/2$, the ground-state consists of $6-$fold degenerate levels. In the presence of a CEF with a tetragonal symmetry, these degenerate levels split into three doublets which are energetically separated from each other. The physical properties of CePd$_2$As$_2$ are greatly influenced by the relative thermal population of these energy levels. In CePd$_2$As$_2$, the magnetic contribution to the entropy, estimated from the specific heat data, reaches $\sim85\%$ of $R\ln2$ at $T_N$ (see Fig.~\ref{HCvsT}). This indicates that the ground-state is a doublet well-separated from the excited CEF levels and that the Kondo effect is rather weak. Further evidence for the localised character of the Ce moments comes from the magnetisation data discussed below.

\begin{figure}[t!]
\centering
\includegraphics[width=1.0\linewidth]{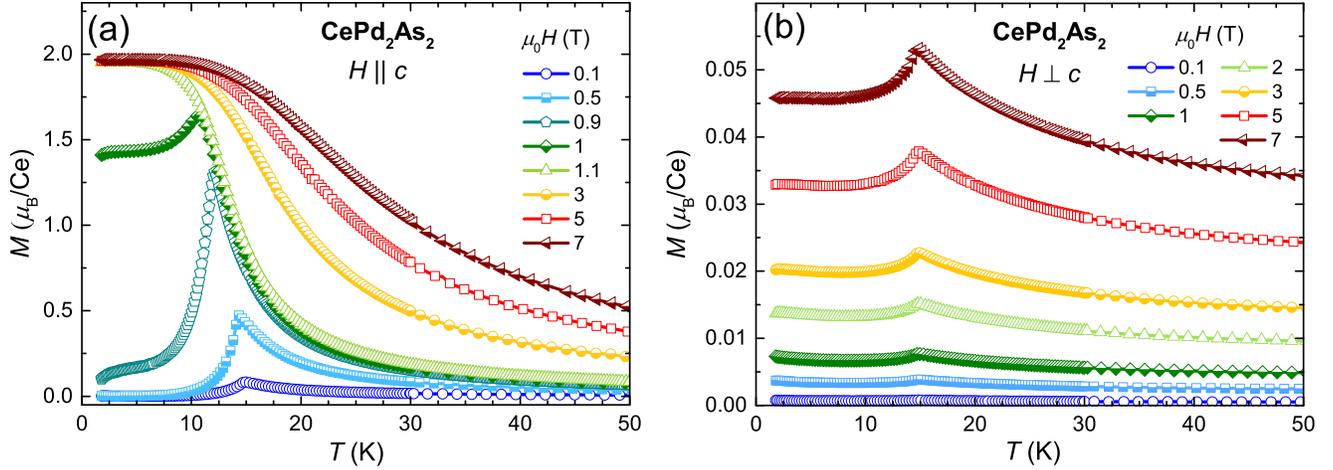}
\caption{Temperature dependence of the magnetisation $M(T)$ measured under various magnetic fields applied (a) parallel and (b) perpendicular to the crystallographic $c$-axis.}
\label{MvsT}
\end{figure}

\begin{figure}[h]
\centering
\includegraphics[width=0.5\linewidth]{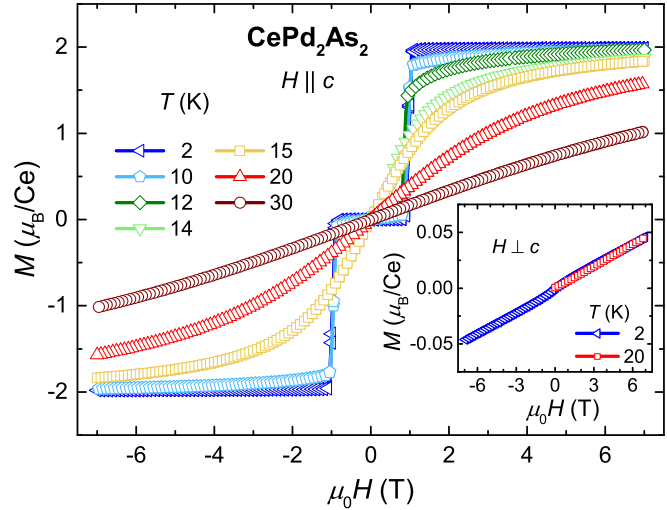}
\caption{Isothermal magnetisation $M(H)$ measured at different temperatures for magnetic fields applied along the $c$-axis and perpendicular to the $c$-axis (inset).}
\label{MvsH}
\end{figure}

\subsection*{Magnetisation}\label{magnetisation}

Figure~\ref{MvsT} presents the temperature dependence of the magnetisation measured under various magnetic fields applied parallel and perpendicular to the crystallographic $c$-axis. For $H\parallel c$, $T_{N}$ shifts to lower temperatures upon increasing the magnetic field, which is expected for an antiferromagnet. As the magnetic field approaches 1~T, the peak in $M(T)$ corresponding to the AFM transition disappears and a broad step-like feature with a saturation of $M(T)$ toward low temperatures develops. However, for $H\perp c$ the position of peak corresponding to $T_{N}$ is independent of the magnetic field and still clearly visible at 7~T. These different behaviours reflect the large magnetic anisotropy present in CePd$_2$As$_2$. We note that, the sudden decrease in the magnetisation below $T_{N}$ for $H\parallel c$ compared to that of $H\perp c$ suggests that the crystallographic $c$-direction is the magnetic easy-axis.

\begin{figure}[t!]
\centering
\includegraphics[width=0.5\linewidth]{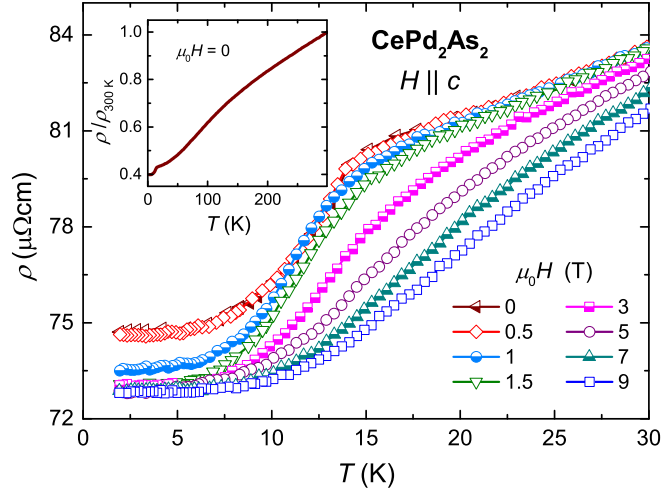}
\caption{Temperature dependence of electrical resistivity $\rho(T)$ of CePd$_2$As$_2$ measured at various magnetic fields applied along the $c$-axis. Inset: normalised resistivity $\rho/\rho_{300~K}$ as a function of $T$.}
\label{rhovsTH}
\end{figure}

\begin{figure}[h!]
\centering
\includegraphics[width=0.85\linewidth]{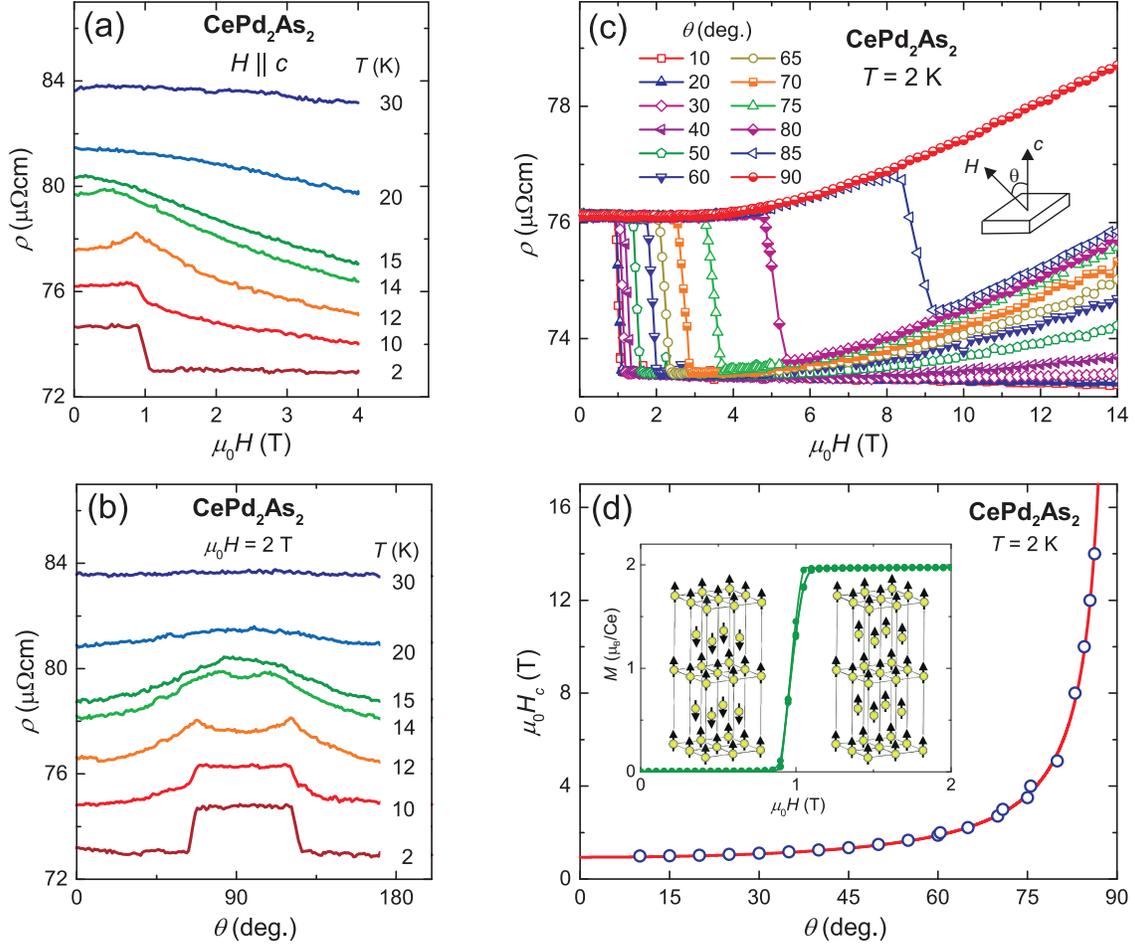}
\caption{(a) Magnetic field dependence of $\rho$ for field parallel to $c$-axis. (b) $\rho$ as a function of the angle ($\theta$) between the magnetic field $H$ and $c$-axis at different temperatures. (c) Magnetic field dependence of $\rho$ for different angles $\theta$ at $T= 2$~K. (d) Variation of metamagnetic critical field $H_c$ as function of $\theta$ at $T = 2$~K. The solid red line is a fit to the data using the equation $H_{c}=H_{c0}/\cos(\theta)$. The inset illustrates the magnetic structures in the A-type antiferromagnetic and in the field-polarised phase.}
\label{transport}
\end{figure}

The isothermal magnetisation $M(H)$ at 2~K, shown in Fig.~\ref{MvsH}, displays a sudden jump at $\mu_0H_{c} \approx 1$~T for $H\parallel c$ followed by an immediate saturation. The observed saturation moment of $2.0~\mu_B/{\rm Ce}$ is in reasonable agreement with the theoretical value of $g_J J=2.14\mu_B$ (where $g_J = 6/7$) expected for a free Ce$^{3+}$ ion. The sudden jump in magnetisation to the saturation value is a typical signature of a spin-flip metamagnetic transition. In the spin-flip process, the spins in the AFM sublattice, which are antiparallel to the field direction, are flipped at $H_c$. Hence, the antiferromagnetism changes to a field-polarised phase in a sudden, single step. The sharp nature of the jump in magnetisation with a small hysteresis point to a first-order type transition.  At higher temperatures, the metamagnetic transition in $M(H)$ broadens and saturates at much higher fields. In the case of  $H\perp c$ (inset of Fig.~\ref{MvsH}), the magnetisation increases monotonously and reaches at 7~T only $2.5\%$ of the saturation value for $H\parallel c$. Furthermore, magnetisation measurements in pulsed fields up to 60~T show a linear increase without any tendency to saturation (not shown). This suggests the absence of any metamagnetic transition for $H\perp c$, which stipulates the huge magnetic anisotropy in CePd$_2$As$_2$.

\subsection*{Electrical transport}
\label{resistivity}

The electrical resistivity $\rho(T)$ of CePd$_2$As$_2$ upon cooling displays a metallic behaviour with a broad curvature at intermediate temperatures, before showing a pronounced kink at about 15~K indicating the AFM transition (inset of Fig.~\ref{rhovsTH}). The broad curvature in the resistivity may be due either to interband scattering or to weak additional spin scattering originating from thermal population of excited CEF levels. At low temperatures, the AFM ordering leads to a sudden decrease in $\rho(T)$ due to the loss of spin-disorder scattering contribution below $T_N$. Figure~\ref{rhovsTH} shows the $\rho(T)$ data recorded at different magnetic fields applied along the $c$-axis. Upon increasing the field up to 1~T, the kink indicating $T_N$ shifts to lower temperatures and becomes washed out, in good agreement with the results from the magnetic susceptibility. Moreover, above 1~T the residual resistivity shows a sudden reduction which coincides with the metamagnetic critical field.

The field and angular dependencies of the resistivity, plotted in Fig.~\ref{transport}(a-c), give further insights into the nature of the metamagnetic transition. At low temperatures, upon increasing the magnetic field $\rho(H)$ suddenly drops at the onset of the metamagnetic transition at the critical field $\mu_0H_c \approx 1$~T. This feature can be attributed to a Fermi-surface reconstruction while going from the AFM to the field-polarised phase. At higher temperatures, the metamagnetic transition in $\rho(H)$ broadens. A small increase in $\rho(H)$ is observed just below $H_c$ in the AFM phase for temperatures close to $T_N$. This could be due to an increased scattering during the spin-flip process associated with the transition from the AFM to field-polarised state\cite{Yamada73}.  Above the AFM transition temperature, $\rho(H)$ displays a gradual decrease upon increasing magnetic field, suggesting a crossover from the paramagnetic to the field-polarised phase. The variation of $\rho$ as function of the angle $(\theta)$ between the magnetic field ($\mu_0H=2$~T) and the crystallographic $c$-axis at different temperatures is shown in Fig.~\ref{transport}(b).  The step-like behaviour at lower temperatures changes to a gradual decrease in resistivity above $T_N$, where the system undergoes a crossover from paramagnetic to the field-polarised phase. Above 30~K, the resistivity becomes independent of the field orientation. Finally, Fig.~\ref{transport}(c) presents the resistivity as a function of field for different angles $\theta$. The metamagnetic critical field $H_c$ increases upon increasing $\theta$ and diverges for $\theta\rightarrow 90^{\circ}$. No drop in $\rho(H)$ is observed up to 14~T for field perpendicular to the $c$-axis. This is consistent with our magnetisation experiments.

\section*{Discussion}

\begin{figure}[t!]
\centering
\includegraphics[width=0.5\linewidth]{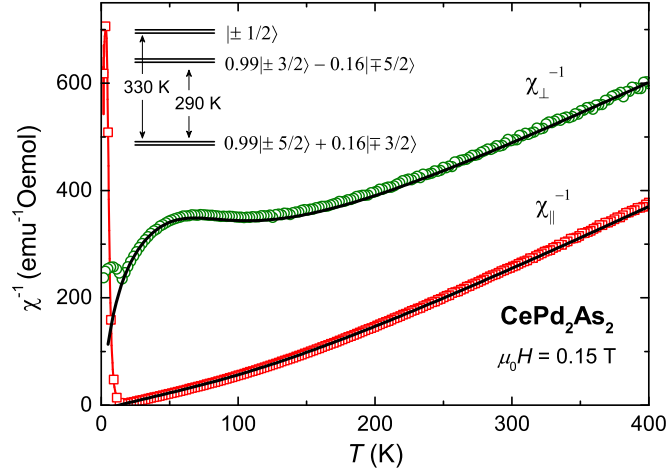}
\caption{Temperature dependence of inverse magnetic susceptibility of CePd$_2$As$_2$. The solid lines are based on CEF calculation. A schematic representation of the CEF levels is shown in the inset.}
\label{CEFfit}
\end{figure}

In CePd$_2$As$_2$, the temperature dependence of the magnetic susceptibility below room temperature strongly deviates from a Curie-Weiss behaviour. This can be attributed to CEF effects. In order to establish the CEF scheme and learn more about the magnetic anisotropy in CePd$_2$As$_2$, we performed a detailed CEF analysis based on our magnetic susceptibility data. For a Ce atom in a tetragonal site symmetry, the CEF Hamiltonian can be written as,
\begin{equation}\label{HCEF}
  \mathcal H_{\mathrm{CEF}} = B^0_2O^0_2 + B^0_4O^0_4 + B^4_4O^4_4,
\end{equation}
where $B^n_m$ and $O^n_m$ are the CEF parameters and the Stevens operators, respectively\cite{Hutchings64, Stevens52}. The magnetic susceptibility including the Van Vleck contribution is calculated as,
\begin{equation}\label{chiCEF}
  \mathcal \chi_{\mathrm{CEF},i} = N_A(g_J\mu_B)^2\frac{1}{Z}\Bigg(\sum\limits_{\substack{m\neq n}} 2\mid \bra{m}J_i\ket{n} \mid ^2 \frac{1-e^{-\beta(E_n-E_m)}}{E_n-E_m}e^{-\beta E_n}+\sum\limits_{\substack{n}} \mid \bra{n}J_i\ket{n} \mid ^2 \beta e^{-\beta E_n}\Bigg),
\end{equation}
where $Z = \sum_{n} e^{-\beta E_n}$, $\beta = 1/k_BT$ and $i=x,y,z$. The inverse magnetic susceptibility including the molecular field contribution $\lambda_i$ is calculated as $\chi_i^{-1}=\chi_{\mathrm{CEF},i}^{-1}-\lambda_i$. $\chi_i^{-1}(T)$ is fitted simultaneously to the experimental data for both field orientations (see Fig.~\ref{CEFfit}).
The data in the paramagnetic phase are well reproduced by the CEF model with a doublet ground-state $\ket*{\Gamma_7^{(1)}} = 0.99\ket{\pm5/2}+0.16\ket{\mp3/2}$ and the excited doublet states $\ket*{\Gamma_7^{(2)}} = 0.99\ket{\pm3/2}-0.16\ket{\mp5/2}$ and  $\ket{\Gamma_6} = \ket{\pm1/2}$ at 290~K and 330~K, respectively. An illustration of the CEF level scheme is shown in the inset of Fig.~\ref{CEFfit}. The crystal field parameters extracted from the model are $B^0_2=-18.66$~K, $B^0_4=-0.22$~K and $B^4_4=1.67$~K, with a molecular field contribution $\lambda_c=-8$~emu$^{-1}$mol along the \emph{c}-axis. It is clear from our CEF analysis that the ground-state is an almost pure $\ket{\pm5/2}$ CEF doublet which is well-separated from the excited doublets. The saturation magnetisation along the $c$-axis for the obtained CEF ground-state is 2.06~$\mu_{\rm B}/{\rm Ce}$, which is in good agreement with the experimental saturation magnetisation of $2.0~\mu_B/{\rm Ce}$. Furthermore, the CEF parameter $B^0_2$ is directly related to the paramagnetic Curie-Weiss temperatures $\theta^{\perp}_{\rm W}$ and $\theta^{\|}_{\rm W}$, along both principal crystallographic directions, as $\theta^{\perp}_{\rm CW}-\theta^{\|}_{\rm CW}=\frac{3}{10}B^0_2(2J-1)(2J+1)$~\cite{Wang71,Bowden71}. Using the experimental values of $\theta^{\perp}_{\rm W}$ and $\theta^{\|}_{\rm W}$, we obtain $B^0_2=-22.5~K$ in good agreement to $B^0_2=-18.66$~K from the CEF-model fit.

Deeper insights into the magnetic structure of the ordered phases in CePd$_2$As$_2$ can be obtained from the magnetisation and electrical resistivity data measured at different orientations of the magnetic field. The magnetisation data suggest that the crystallographic $c$-direction is the easy-axis of the magnetisation. Moreover, the small magnetisation in the $ab$-plane compared with the large magnetisation along the $c$-axis indicates an AFM structure with the spins pointing along the $c$-axis. In addition, the spins are locked along the $c$-axis by the magneto-crystalline anisotropy, as indicated by the absence of a metamagnetic transition for magnetic field up to 60~T applied perpendicular to the $c$-axis. These observations confirm that the moments in CePd$_2$As$_2$ are Ising-type. The Ising-nature of the spins is also supported by the angular dependence of the metamagnetic critical field extracted from the electrical resistivity data shown in Fig.~\ref{transport}(c). The resulting angular dependence of $H_c$ is displayed in Fig.~\ref{transport}(d). $H_c(\theta)$ increases sharply for $\theta\rightarrow90^{\circ}$ and $H_c$ is not detected for field oriented perpendicular to the $c$-axis.

In order to understand the angular dependence of $H_c(\theta)$, we fit the data by the equation,
\begin{equation}\label{BcvsTheta}
  H_c = \frac{H_{c0}}{\cos(\theta)}
\end{equation}
where $H_{c0}$ is the critical field for field parallel to the $c$-axis. Equation \ref{BcvsTheta} describes the experimental data very well with $\mu_0H_{c0} = 0.95$~T. In other words, the metamagnetic transition occurs only when the component of magnetic field along the $c$-axis reaches the value of $H_{c0}$. Based on these results, we can conclude the following scenario for the spin structure of CePd$_2$As$_2$: in the AFM phase, the Ce moments are aligned along the $c$-axis and are locked along this axis by the magneto-crystalline anisotropy. When the component of external magnetic field along the $c$-axis exceeds $H_{c0}$, the anti-parallel spins undergo a spin-flip transition to the field-polarised ferromagnetically ordered phase.

\begin{figure}[t!]
\centering
\includegraphics[width=0.7\linewidth]{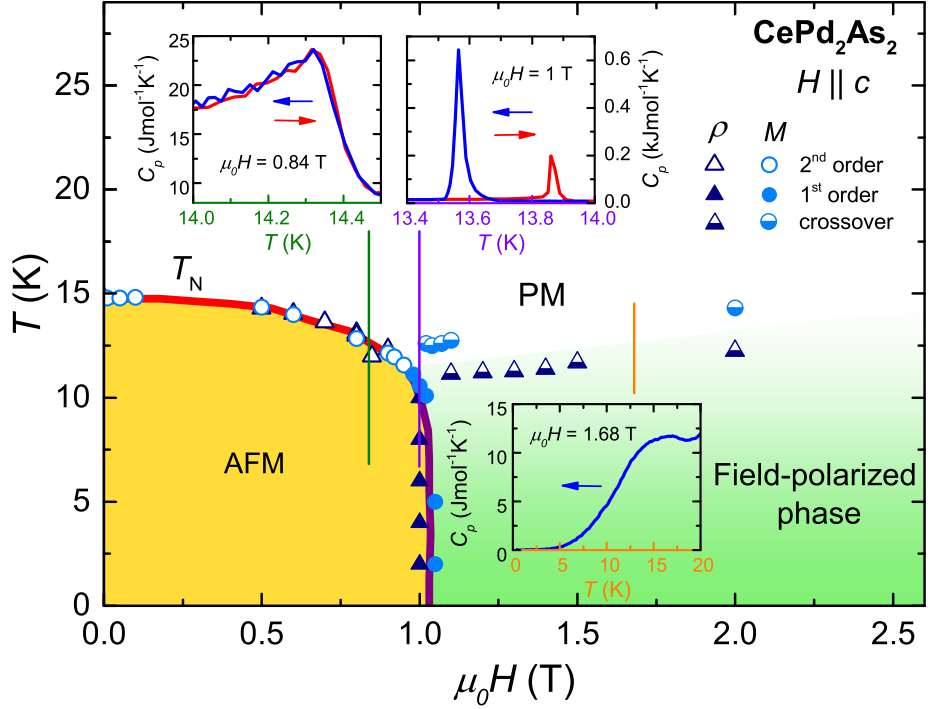}
\caption{Magnetic phase diagram of CePd$_2$As$_2$ for $H\parallel c$ summarising the results from magnetisation and electrical resistivity data. The lines are guides to the eyes. The three insets present the temperature dependence of the specific heat at three representative magnetic fields as indicated by the vertical lines. The red and blue arrows in the insets indicate the direction of the temperature sweep.}
\label{THphaseD}
\end{figure}

The single, sharp jump in the magnetisation with a weak hysteresis at the first-order metamagnetic transition from the AFM to the ferromagnetically polarised state points at a simple spin structure of the AFM phase. The small value of the critical field $H_{c0}$, compared to the value of $T_N$, indicates that, in terms of a Heisenberg model with a few different inter-site exchange interactions, the AFM ones are much weaker than the ferromagnetic (FM) ones. Because of the topology of the tetragonal body centered Ce sublattice, a strong FM interaction between atoms in adjacent layers competing with a weak in-plane AFM interaction would always result in a FM ground-state. In contrast, a strong FM in-plane interaction with a weak AFM inter-plane interaction can easily account for all observations. In addition, we note that isovalent substitution of P for As results in a FM ground-state~\cite{Shang14}. Thus all these properties provide strong indication that the AFM structure of CePd$_2$As$_2$ is just a simple AFM stacking of FM layers. Substituting P for As is just turning the inter-plane exchange from weakly AFM to FM. Therefore, we propose a magnetic structure with weakly antiferromagnetically coupled FM layers of Ising-spins in the AFM state of CePd$_2$As$_2$, as illustrated in the inset of Fig.~\ref{transport}(d). A mean-field approximation based on a two-sublattice model can appropriately describe such a spin system. According to this model, the spin-flip occurs when the applied magnetic field is able to overcome the inter-layer AFM coupling. Therefore, the metamagnetic critical field can be expressed as $H_c = \lambda_{\rm AFM} M $, where $\lambda_{\rm AFM}$ is the inter-sublattice molecular field constant and $M$ is the magnetisation of the ferromagnetic state\cite{Boer03}. Similarly, the intra-sublattice molecular field constant $\lambda_{\rm FM}$ can be extracted from the relation $T_N = \frac{1}{2} C(\lambda_{\rm FM}-\lambda_{\rm AFM})$, where $C$ is the Curie constant $C = N_A\mu_{\rm 0}g_J^2J(J+1)\mu_{B}^2/3k_{\rm B}$. By using the experimentally obtained values $H_{c0}$, $M_S$ and $T_N$, the inter-layer AFM exchange strength ($z_{\rm AFM}J_{\rm AFM}$) and intra-layer FM exchange strength ($z_{\rm FM}J_{\rm FM}$) are calculated as $-0.25$~K and 9.83~K, respectively. Here, $z_{\rm AFM}$ and $z_{\rm FM}$ are the number of nearest-neighbour spins participating in the respective interactions. The large intra-layer FM exchange strength is consistent with the experimental observations and plays a crucial role in the first-order nature of the metamagnetic transition.

The $T-H$ phase diagram of CePd$_2$As$_2$ for $H\parallel c$, presented in  Fig.~\ref{THphaseD}, summarises our results. At low temperatures, application of a magnetic field induces a metamagnetic transition at $\mu_0H_{c0}=0.95$~T resulting in a field-polarised phase. Above $T_N$, CePd$_2$As$_2$ shows a crossover behaviour from the paramagnetic to the field-polarised phase, reflected by the broad features in magnetisation and electrical resistivity. Additional information on the nature of the transitions between the various phases can be obtained from specific heat data. The temperature dependencies of the specific heat $C_p$ of CePd$_2$As$_2$ for three representative magnetic fields are plotted in insets of Fig.~\ref{THphaseD}. A cusp in $C_p(T)$ indicates the transition form the paramagnetic to AFM phase at 0.84~T. The second order nature of this transition is evidenced by the absence of any thermal hysteresis in the data. In contrast, a strong thermal hysteresis and a spike-like anomaly in $C_p(T)$ at 1~T confirms the first-order nature of the metamagnetic transition from the AFM to the field-polarised phase. Finally, the crossover from the paramagnetic to the field-polarised phase at higher magnetic fields is reflected by a broad, hump-like feature in $C_p(T)$.

\par
Because of its strong Ising anisotropy and simple magnetic behaviour, CePd$_2$As$_2$ is a nice example to illustrate a misinterpretation frequently encountered in the analysis and discussion of magnetic properties of Ce- and Yb-based compounds. The Weiss temperatures determined from Curie-Weiss fits to the high temperature part of the susceptibility are frequently argued to reflect the anisotropy, the sign and the magnitude of the exchange interactions. In CePd$_2$As$_2$ this would lead to the conclusion that the exchange in the basal plane is strongly antiferromagnetic while the exchange along the \emph{c}-direction is weaker and ferromagnetic. Our analysis clearly demonstrates that this conclusion would be completely wrong, because the Weiss temperatures determined from Curie-Weiss fits at high temperatures are dominated by the effect of the CEF. Except for special cases, CEF generally result in a seemingly AFM, negative $\theta_{\rm W}$ for the direction of the small CEF ground-state moment and an apparently FM, positive $\theta_{\rm W}$ for the direction of the large CEF ground-state moment.

\section*{Summary}\label{summary}

We have investigated the magnetic properties and the CEF scheme of CePd$_2$As$_2$ by detailed temperature, magnetic field and angular dependent magnetic, thermodynamic and electrical transport studies on single crystalline samples. The detailed CEF analysis based on the magnetic-susceptibility data indicates an almost pure $\ket{\pm5/2}$ CEF ground-state doublet with the dominantly $\ket{\pm3/2}$ and the $\ket{\pm1/2}$ doublets at 290~K and 330~K, respectively. CePd$_2$As$_2$ orders antiferromagnetically in a simple A-type order below $T_N=14.7$~K. Our results imply a uniaxial AFM structure with spins locked along crystallographic $c$-axis. An external magnetic field applied along the $c$-axis induces a metamagnetic spin-flip transition at $\mu_0H_{c0}=0.95$~T leading to a ferromagnetic spin alignment. No metamagnetic transition is observed for a magnetic field perpendicular to the $c$-axis, proving the huge Ising-like anisotropy in CePd$_2$As$_2$.

\section*{Methods}

Single crystals of CePd$_2$As$_2$ were synthesised by a self-flux method. Initially, polycrystalline CePd$_2$As$_2$ was obtained by a solid-state reaction as reported previously~\cite{Shang14}. Then the polycrystalline pellet was loaded into an alumina crucible and sealed in an evacuated quartz ampule. The ampule was heated up to $1160^{\circ}$C and held at this temperature for 24 hours, followed by slow cooling to $900^{\circ}$C at the rate of $1.6^{\circ}$C/h. Shiny plate-like single crystals of CePd$_2$As$_2$ were obtained. The crystal orientation and chemical homogeneity were checked by x-ray diffraction (XRD) and energy dispersive x-ray analysis (EDX), respectively. XRD measurements were carried out on a PANalytical X'pert MRD diffractometer with Cu $K_{\alpha1}$ radiation and a graphite monochromator. Magnetisation measurements were carried out in the temperature range $1.8~{\rm K}-400$~K and in magnetic field up to 7~T using a SQUID-VSM (MPMS3, Quantum Design). High-field magnetisation measurements up to 60~T in pulsed magnetic fields were performed at the Dresden High Magnetic Field Laboratory, Germany. The electrical transport experiments were carried out in the temperature range $2~{\rm K}-300$~K and magnetic fields up to 14~T using a Physical Property Measurement System (PPMS, Quantum Design). The electrical resistivity was measured using a standard four-terminal method, where electrical contacts to the sample were made using 25~$\mu$m gold wires and silver paint. The temperature dependence of specific heat was also measured using a PPMS.

\bibliography{CePd2As2}

\section*{Acknowledgements}

We thank Yurii Skoursky for his support in conducting the pulsed-field magnetisation measurement.
This work was partly supported by Deutsche Forschungsgemeinschaft (DFG) through the Research Training Group GRK 1621. The work at Zhejiang University was supported by the National Natural
Science Foundation of China (U1632275), National Key Research and Development Program of China
(No. 2016YFA0300202) and the Science Challenge Project of China.
RDdR acknowledges financial support from the Brazilian agency CNPq (Brazil).

\end{document}